\documentclass[11pt]{article}
\usepackage{epsfig,cite}
\usepackage{graphicx}
\usepackage{multicol}
\usepackage{color}
\usepackage{amssymb}

\def\bea{\begin{eqnarray}}
\def\eea{\end{eqnarray}}

\newcommand{\be}{\begin{equation}}
\newcommand{\ee}{\end{equation}}
\newcommand{\ba}{\begin{array}}
\newcommand{\ea}{\end{array}}

\newcommand\prd[3]   %{\@spires{PHRVA%2CD#1%2C#3}
        {{Phys.\ Rev.\ }{\bf D #1} (#2) #3}
\newcommand\prl[3]   %{\@spires{PRLTA%2C#1%2C#3}
        {{Phys.\ Rev.\ Lett.\ }{\bf #1} (#2) #3}
\newcommand\plb[3]   %{\@spires{PHLTA%2CB#1%2C#3}
        {{Phys.\ Lett.\ }{\bf B #1} (#2) #3}
\newcommand\npb[3]    %{\@spires{NUPHA%2CB#1%2C#3}
        {{Nucl.\ Phys.\ }{\bf B #1} (#2) #3}
\newcommand\app[3]   %{\@spires{APHYE%2C#1%2C#3}
        {{Astropart.\ Phys.\ }{\bf #1} (#2) #3}
\newcommand\jhep[3]  %{\href{http://jhep.sissa.it/stdsearch?paper=#1%28#2%29#3}
        {{J. High Energy Phys.\ }{\bf #1} (#2) #3}
\newcommand\epjc[3]  %{\@spires{EPHJA%2CC#1%2C#3}
        {{Eur.\ Phys.\ J. }{\bf C #1} (#2) #3}
\newcommand\npps[3]  %{\@spires{NUPHZ%2C#1%2C#3}
        {{Nucl.\ Phys.\ }{\bf #1} {\it(Proc.\ Suppl.)} (#2) #3}
\newcommand\jcap[3]  %{\href{http://jhep.sissa.it/stdsearch?paper=#1%28#2%29#3}
        {{JCAP\ }{\bf #1} (#2) #3}

\newcommand\apj[3]    %{\@spires{ASJOA%2C#1%2C#3}
        {{\it Astrophys.\ J. }{\bf #1} (#2) #3}

\newcommand{\gsim}{~{}_{\textstyle\sim}^{\textstyle >}~}
\newcommand{\lsim}{~{}_{\textstyle\sim}^{\textstyle <}~}

\begin{document}
\begin{titlepage}
\pagestyle{empty}
\baselineskip=21pt
\vspace*{0.5cm}
%\rightlineenter here the preprint number}
%\vskip 0.7in
\begin{center}
\hfill {NPAC-09-12}

\vspace*{0.5cm}
{\huge\sf MSSM Baryogenesis and Electric Dipole Moments:\\[0.3cm]
An Update on the Phenomenology
}\\[0.2cm]
\end{center}
\begin{center}
\vskip 0.1in
{\Large\sf  V.~Cirigliano${}^{(a)}$, Yingchuan Li${}^{(b)}$, S.~Profumo${}^{(c)}$ and M.~J.~Ramsey-Musolf${}^{(b)(d)}$}\\
\vskip 0.1in
{\it {(a) Theoretical Division, Los Alamos National Laboratory, Los Alamos NM 87545, USA}}\\
{\it {(b) Department of Physics, University of Wisconsin, Madison, WI 53705, USA}}\\
{\it {(c) Department of Physics and Santa Cruz Institute for Particle Physics\\
University of California, 1156 High St., Santa Cruz, CA 95064,
USA}}\\
{\it {(d) Kellogg Radiation Laboratory, California Institute of
Technology, Pasadena, CA 91125 USA
}}\\[0.2cm]
\mbox{\footnotesize E-mail: {\tt cirigliano@lanl.gov, yli@physics.wisc.edu, profumo@scipp.ucsc.edu, mjrm@physics.wisc.edu}}\\

Draft of \today

\vskip 0.4in
{\bf Abstract}
\end{center}
\baselineskip=18pt \noindent
%%%%%%%%%%%%%%%%%%%%%%%%%%%%%%%%%%%%%%%%%%%%%%%%%%%%%%%%%%%%%%%%%%%%%

\noindent We explore the implications of electroweak baryogenesis
for future searches for permanent electric dipole moments in the
context of the minimal supersymmetric extension of the Standard
Model (MSSM). From a cosmological standpoint, we point out that
regions of parameter space that over-produce relic lightest
supersymmetric particles can be salvaged only by assuming a dilution
of the particle relic density that makes it compatible with the dark
matter density: this dilution must occur after dark matter
freeze-out, which ordinarily takes place after electroweak
baryogenesis, implying the same degree of dilution for the generated
baryon number density as well. We expand on previous studies on the
viable MSSM regions for baryogenesis, exploring for the first time
an orthogonal slice of the relevant parameter space, namely the
($\tan\beta,m_A$) plane, and the case of non-universal relative
gaugino-higgsino CP violating phases. The main result of our study
is that in all cases lower limits on the size of the electric dipole
moments exist, and are typically on the same order, or above, the
expected sensitivity of the next generation of experimental
searches, implying that MSSM electroweak baryogenesis will be soon
conclusively tested.

%%%%%%%%%%%%%%%%%%%%%%%%%%%%%%%%%%%%%%%%%%%%%%%%%%%%%%%%%%%%%%%%%%%%%
\vfill
\end{titlepage}
%\baselineskip=18pt

%\noindent\rule\textwidth{.1pt}
%\tableofcontents
%\vspace*{0.5cm}
%\noindent\rule\textwidth{.1pt}
%\vspace*{0.5cm}

%******************************************************************************
%******************************************************************************
\section{Introduction}
Explaining the origin of the baryonic matter of the Universe remains an open problem at the interface of cosmology with particle and nuclear physics. Consistent values for the ratio $Y_B$ of the baryon number density ($n_B$) to entropy density ($s$) have been obtained from measured light element abundances in the context of Big Bang Nucleosynthesis (BBN) \cite{Yao:2006px}  and acoustic oscillations in the cosmic microwave background (CMB) as measured by the Wilkinson Microwave Anisotropy Probe (WMAP) \cite{Yao:2006px, Dunkley:2008ie}:

\be
Y_B=n_B/s = \left\{
\begin{array}{ccc}
(6.7  \; - \;  9.2) \times 10^{-11}  & \; & {\rm BBN} \; \\%\cite{Yao:2006px} \\
(8.36  \; - \; 9.32) \times 10^{-11}  & \; & \textrm{CMB} \\%\; \cite{Yao:2006px, Dunkley:2008ie}
\end{array} \right.
\ee

Assuming that the $Y_B=0$ at the end of the inflationary epoch, the dynamics of the subsequently evolving cosmos would then have generated a non-vanishing baryon asymmetry. As first noted by Sakharov, \cite{Sakharov:1967dj}, these dynamics must have included violation of baryon number (B) conservation; violation of C and CP conservation; and a departure from equilibrium dynamics assuming CPT invariance.

Although it is not known when in the history of the universe these ingredients came into play, one possibility that can be tested with terrestrial experiments is that baryogenesis occurred during the era of electroweak symmetry breaking. In this scenario, which we consider here, a departure from equilibrium occurs through a strong, first order electroweak phase transition (EWPT) in which bubbles of broken electroweak symmetry form in the symmetric background. Charge asymmetries generated by CP-violating interactions at the bubble walls diffuse into the unbroken phase, where B-violating electroweak sphalerons convert them into non-vanishing baryon number density. The expanding bubbles capture this baryon number density by quenching the sphaleron transitions, ultimately leading to a relic baryon asymmetry at zero temperature.

In principle, the Standard Model (SM) of particle physics could have satisfied these \lq\lq Sakharov criteria" as needed for electroweak baryogenesis, particularly as the B-violating sphaleron transitions associated with the SU(2$)_L$ gauge sector are unsuppressed at temperatures of order 100 GeV and above. However, the strength of the SM CP-violating (CPV) interactions are too small to have generated sufficiently large charge asymmetries needed to bias the sphalerons into making baryons. In addition, the lower bound on the mass of the SM Higgs boson \cite{leph} implies that finite temperature SM effective scalar potential that governs electroweak symmetry-breaking  would not admit a sufficiently strong first order EWPT as needed to prevent a \lq\lq washout" of any baryon asymmetry that might otherwise have been produced at high temperatures \cite{noewb ,Bochkarev:1987wf}. Consequently, an explanation of the observed $Y_B$ requires new physics beyond the Standard Model .

In this paper, we concentrate on the possibility that this new physics involved TeV scale supersymmetric interactions in the guise of the minimal supersymmetric Standard Model (MSSM). The general motivation for considering TeV scale supersymmetry is well-known, and we refer the reader to the vast literature for a discussion \cite{baerbook}. Here, we focus on the minimal scenario, up-dating our previous study to take into account several developments (outlined below). The phenomenological constraints on MSSM electroweak baryogenesis (EWB) are quite stringent: the existence of a strong first order EWPT requires both a relatively light SM-like Higgs scalar and a correspondingly light scalar superpartner of the right-handed top quark. Searches for scalars at the Large Hadron Collider (LHC) will readily explore the narrow window of available parameter space for these considerations. The strengths of MSSM CPV interactions relevant for EWB are similarly tightly bounded by current limits on the permanent electric dipole moments (EDMs) of the electron, neutron, and mercury atom, with bounds on the latter having recently improved by a factor of seven \cite{Griffith:2009zz}. Thus, it is conceivable that the combination of LHC studies and planned EDM searches having substantially improved sensitivities may rule out the minimal version of supersymmetric EWB, leaving other scenarios such as non-minimal supersymmetry or leptogenesis as the most viable possibilities. The purpose of the present study is to delineate just how close we are to such a situation.

The new recent developments in supersymmetric EWB include the following results:
\begin{itemize}
\item[(i)] Carena {\em et al.} \cite{Carena:2008rt} have completed an analysis of the EWPT in the MSSM using a two-loop, finite temperature effective potential computed with effective theory to integrate out all but the most relevant scalar degrees of freedom. Their results indicate that a sufficiently strong first order EWPT can occur if the SM-like Higgs and RH stop masses are less than $~125$ GeV, with some dependence on the cutoff scale of the effective theory. For the EWPT-viable regions of parameter space, the electroweak minimum is metastable with a lifetime  longer than the age of the universe. The presence of the light, RH-stop leads to the existence of a deeper zero-temperature color-breaking minimum that can be avoided if the electroweak minimum is deeper at temperatures associated with the EWPT.
\item[(ii)] Three of us have recently completed a two-loop computation of the electric dipole moments of the electron and neutron in the MSSM, including all of the CPV interactions that are most likely to be responsible for EWB in the MSSM \cite{Li:2008kz}. The results indicate that even under the most optimistic scenarios, viable MSSM EWB requires that the CPV phases in the gaugino-Higgs-Higgsino sector be non-universal, in contrast to the usual universality assumptions. The CPV interaction most likely to lead to successful EWB involves the bino and Higgsino fermions \cite{Li:2008ez}. 
\item[(iii)] It was shown in Ref. \cite{Chung:2009qs} that the value of $\tan\beta$ could play a decisive role in the transport dynamics that convert the bino-driven Higgsino asymmetry into a left-handed SM fermion number asymmety -- the quantity that ultimately biases the sphalerons into making baryons. Larger values of $\tan\beta$ lead to a suppression of $Y_B$ and a potentially significant dependence on the masses of the RH bottom squarks (${\tilde b}_R$) and tau sleptons (${\tilde\tau}_R$), whose Yukawa couplings are enhanced with increasing $\tan\beta$. For values of this parameter favored by the muon anomalous magnetic moment, one expects a suppression of a factor of two or more -- and possibly even a sign change in $Y_B$ -- depending on the RH sbottom and stau masses.
\end{itemize}

In addition to analyzing the phenomenological implications of some of these
developments for MSSM EWB and EDM searches, we also take into
account another physical consideration not treated in previous work,
namely the impact of the cold dark matter (CDM) relic density at the
time the baryons are produced. In some regions of parameter space in
which MSSM EWB is na\"\i vely viable, one expects
$\Omega_\mathrm{CDM}$ due to neutralinos to be larger than the
observed relic density. In standard cosmological scenarios, these
parameter space regions are therefore excluded. They may become
viable in non-standard scenarios  involving late-time reheating
that lead to a dilution of the neutralino density due to entropy
production. This entropy dilution, however, will also lead to a
reduction in $Y_B$. We find that the resulting impact on the
EWB-allowed MSSM parameter space can be substantial, particularly in
regions where wino-Higgsino driven EWB might otherwise have been
potentially important. The impact on the bino-Higgsino driven
scenario is less pronounced since in this region of parameter space
the neutralino relic density typically lies below the observed value
so that no entropy dilution rescaling occurs.

Taking all of the aforementioned considerations into account, we
find minimum values for $| d_e|$ and $|d_n|$ for which MSSM EWB
would be viable under the most optimistic scenarios. These values
lie somewhat below the expected sensitivities of the next generation
of EDM searches, though given outstanding theoretical uncertainties
(discussed below) we cannot make a definitive statement until further
theoretical progress is achieved. For the interim, we provide this
analysis as a phenomenological \lq\lq progress report" that will
undoubtedly require further updates in the future. To this end, we
illustrate the key features of the current status in several
figures: (a) a series (Fig. 1 - 6) that shows the EWB-viable regions
in  the  plane of gaugino-Higgsino mass parameters   for different values of
the CP-violating phases and corresponding values of the EDMs. In
this series, we show the impact of imposing the requirements of
entropy dilution needed for consistency with the observed CDM relic
density; (b) a series of plots (Fig. 7 - 10) in the space of
$\tan\beta$ and the mass of the CP-odd scalar, $m_A$.

We discuss in Section \ref{sec:setup} the general framework, and we outline the MSSM parameter space compatible with EWB. In this section only we consider a light right handed stop as the driver of a strongly first order EW phase transition: in the rest of the paper we drop this assumption, and always consider (i) a heavy sfermion sector (including both stops) and (ii) beyond the MSSM physics driving a strongly first order EWPT. In Section \ref{sec:oh2} we discuss the
relation between the neutralino relic abundance and EWB . For
purposes of comparison with our earlier work, we consider here a
scenario of universal CPV gaugino-Higgsino-Higgs phases without
imposing the EDM constraints. The latter are treated in Sections
\ref{sec:edm} and \ref{sec:bino}. Constraints in the
$\tan\beta$-$m_A$ plane are given in Section \ref{sec:tbma}. We
summarize our findings and the outstanding theoretical issues in
Section \ref{sec:conclusions}.

\section{MSSM Baryogenesis and EDMs}\label{sec:setup}

For purposes of our analysis, in this section (and only here) we start with the customary assumption that the RH stop and SM-like Higgs masses lie in the ranges indicated by the analysis of Ref. \cite{Carena:2008vj}, so that the universe undergoes a sufficiently strong first order EWPT to prevent baryon number washout (we relax this assumption and consider a heavy sfermion sector, including both stops, everywhere else in this analysis). Consequently, we consider a scenario under which the leading CPV particle asymmetry generation occurs in the gaugino-Higgsino sector. In order to accommodate the lower bound on the lightest SM-like Higgs boson, the LH stop mass must be of order one TeV, thereby leading to Boltzmann suppression of LH stops and the production of large CPV asymmetries in the stop sector. In principle, one could counteract this suppression to some extent through resonant CPV stop processes wherein $m_{{\tilde t}_L}\sim m_{{\tilde t}_R}$, but our assumption of a heavy sfermion sector suppresses the importance of this contribution, even if resonant. For relatively large $\tan\beta$ leading to enhanced bottom and tau Yukawa couplings, one might introduce additional sources associated with these scalar quarks, but we leave consideration of this possibility to a future study. The remaining source of CPV asymmetries thus arises in the gaugino-Higgsino sector, for which one may consider relatively light superpartner masses (of order a few hundred GeV).  The corresponding CPV phases are, after appropriate field redfinitions:
\be
\label{eq:cpvphase}
\phi_j = \mathrm{Arg}(\mu M_j b^\ast)\ \ \ ,
\ee
where $\mu$ is the supersymmetric Higgsino-Higgs mass parmeter; $M_j$ are the SUSY-breaking soft mass parameters for the bino ($j=1$) and wino ($j=2$); and $b$ is a soft SUSY-breaking Higgs mass parameter.

To obtain $Y_B$ as a function of the MSSM parameters, we rely on the
work of Ref.~\cite{Lee:2004we,Cirigliano:2006wh,Chung:2008aya,Chung:2009qs},
which include the parameter dependence of both the CPV source terms
in the quantum Boltzmann equations as well as that of the
CP-conserving, particle number-changing interactions. Both sets of
effects are required for a robust prediction: the CPV source terms
determine the overall scale of CPV Higgsino-Higgs asymmetries that
may arise, while the particle number-changing interactions determine
the efficiency with which these asymmetries convert into the
left-handed fermion density that biases the electroweak sphalerons.
To lowest order in Yukawa and gauge couplings, particle number
changing processes occur through decays and inverse decays 
$A\leftrightarrow B+C$; in regions where these reactions are
kinematically forbidden, $A+B\leftrightarrow C+D$ scattering
reactions dominate. We avoid these parameter space regions as there
does not yet exist a complete study of the parameter dependence of
these scattering rates.

For the CPV sources, we also rely on the computation of Ref.
\cite{Lee:2004we}, which computed the sources to leading non-trivial
order in the spacetime varying Higgs background field. This
approximation is likely to overestimate the source strength
\footnote{See, e.g. Refs
\cite{Carena:2002ss,Konstandin:2004gy,Konstandin:2003dx}}, so we
consider our results for $Y_B$ to represent the most optimistic
scenario, and therefore, the most conservative for statements about
the impact of EDMs. On-going theoretical work is aimed at resumming the
background field while consistently including the effects of
diffusion ahead of the bubble wall and interactions that would drive
the out-of-equilibrium plasma toward equilibrium. Once such a
consistent framework has been achieved, our analysis will likely
need to be up-dated.

We also note that the sources of Ref. \cite{Lee:2004we} apply to
only resonant production of CPV Higgsino asymmetries, and we do not
consider non-resonant processes that generally require substantially
larger CPV phases in order to produce the observed $Y_B$ (see, e.g.
Ref. \cite{Carena:2002ss}). Resonant EWB occurs when the mass
parameters of two particles having a trilinear coupling with the
Higgs scalar are nearly degenerate. In the case of top squarks, for
example, resonant asymmetry production occurs when for nearly equal, and light enough
LH and RH stop masses, a situation precluded by the requirements of
a strong first order EWPT and the bound on the lightest Higgs mass
as noted above. For the gauginos and Higgsinos, resonant EWB arises
when $\mu\sim M_{1,2}$ -- the scenario on which we concentrate
here\footnote{In our conventions, the Higgs and Higgsino fields have
been rotated so that $\mu$ is real and positive}.

Within this context, the overall source strength depends crucially
on both the CPV phases and $\Delta\beta$, the change in $\tan^{-1}
v_u/v_d$ from the bubble wall exterior to its interior with $v_u$
and $v_d$ being the spacetime-dependent vacuum expectation values
(or background fields) associated with the neutral up- and down-type
Higgs scalars, respectively. The authors of Ref.
\cite{Moreno:1998bq} showed that $\Delta\beta$ decreases with
increasing $m_A$, and we take this $m_A$-dependence into account in
our analysis of the parameter space. Doing so is not entirely
consistent with the phase transition analysis of Ref.
\cite{Carena:2008vj}, which considered $m_A$ larger than $\sim 10$
TeV. We expect  that a two-loop EWPT analysis allowing for a
dynamically active, light CP-odd Higgs would not change the allowed
ranges of $m_{{\tilde t}_R}$ and $m_h$ substantially, and we
therefore consider values of $m_A$ below one TeV as needed for
unsuppressed, resonant, CPV sources.

The dependence of $Y_B$ on $\tan\beta$  remains a topic of on-going
research. To our knowledge, there exist no results in the
literature, for the $\tan\beta$-dependence of $\Delta\beta$. On the other hand, a previously
overlooked $\tan\beta$-dependence of transport dynamics has been
observed in Refs. \cite{Chung:2009qs}. Earlier work had assumed that
the CP-conserving, particle number changing processes were dominated
by stop quark (squark) Yukawa interactions, given the large value of
the top Yukawa coupling relative to that of the other (s)fermions.
For moderate $\tan\beta$, however, the (s)bottom and (s)tau Yukawa
couplings are relatively enhanced, and their impact on the
conversion of a Higgs-Higgsino asymmetry into the net left-handed
fermion density ($n_\mathrm{left}$) that couples to electroweak
sphalerons can be substantial. In what follows, we include this
$\tan\beta$- dependence by re-scaling $Y_B$ according to Fig. 1 of
Ref. \cite{Chung:2008aya}, that shows the ratio of $Y_B$ computed
including the bottom and tau Yukawa couplings to the baryon
asymmetry computed in their absence as a function of $\tan\beta$.
For purposes of this analysis, we work in the limit of heavy
${\tilde\tau}_R$ , but note that -- as shown in Fig. 1 of Ref.
\cite{Chung:2008aya} -- the presence of a light ${\tilde\tau}_R$ can
lead to a suppression of $Y_B$. We also work in the region of
parameter space wherein $m_{{\tilde b}_R} >> m_{{\tilde t}_R}$ to
avoid further suppression of $Y_B$ or even a sign change that can
occur for $m_{{\tilde b}_R} \lsim m_{{\tilde t}_R}$.

For the computation of bounds from EDMs, we draw on the two-loop
computations of Ref. \cite{Li:2008kz}. Limits from one-loop EDMs
generally imply values of $|\sin\phi_j|$ that are too small to
accommodate the observed value of $Y_B$, even in the regions of
resonant EWB. To circumvent these bounds, we work in the limit of
heavy first generation sfermions, thereby suppressing one-loop EDMs
without quenching $Y_B$ via Boltzmann suppression of stops,
gauginos, and Higgsinos. In this limit, both the elementary fermion
EDMs and the chromo-EDMs of quarks are suppressed. As a result,
constraints from the $^{199}$Hg atomic EDM limit are minimal, as the
$^{199}$Hg atomic EDM is dominated by the chromo-EDMs in the MSSM.
In contrast, significant bounds from the two-loop EDMs of the
electron and quarks (contributing to $d_n$) can occur, with the most
stringent arising from graphs involving closed chargino and
neutralino loops and the exchange of two bosons as in Fig. 1 of Ref.
\cite{Li:2008kz}. To analyze these constraints, we rely on the work
of Ref. \cite{Li:2008kz} and consider two cases: (a) universal
gaugino phases, wherein $\phi_1=\phi_2$, and (b) non-universal
phases, for which $\phi_1$ and $\phi_2$ are allowed to differ. For
the second scenario, the sensitivity of the electron and neutron
EDMs to $\phi_2$ is roughly fifty times stronger than the
sensitivity to $\phi_1$.

In summary, the most optimistic scenario leaving the largest window
for viable MSSM EWB occurs in the following region of parameter
space: a light right-handed stop with $m_{{\tilde t}_R}\lsim 125$
GeV; the remaining sfermion masses being heavy, $m_{\tilde f}\gsim
1$ TeV; light gauginos and Higgsinos with $\mu\sim M_1$;
non-universal phases: $\phi_1\not=\phi_2$; a relatively light
pseudoscalar Higgs and moderate to small $\tan\beta$. In what
follows, we illustrate the impact of relaxing these assumptions.

The MSSM parameter space compatible with EWB is severely constrained by the requirement that the lightest neutralino be the lightest supersymmetric particle (LSP), since a light RH stop with $m_{{\tilde t}_R}\lesssim 125$ GeV forces the LSP mass to also be below that value\footnote{A caveat is the assumption of a gravitino LSP, but we do not consider this possibility here.}. However, the requirement of a light RH stop can be relaxed in several minimal extensions to the MSSM that include for instance an extended Higgs sector. One large class of such models are those that include a gauge singlet with couplings to the SU(2) Higgs doublets \cite{singlet}. Other viable beyond-the-MSSM setups featuring a strongly first order EWPT also include models with a fourth fermion generation \cite{Fok:2008yg} and top-flavor models \cite{Shu:2006mm}; models with higher dimensional scalar representations of SU(2) (which might lead to a multistep EWPT \cite{mjrminprep}) provide yet another promising such framework \cite{su2}. In several of these models, the physical mechanism that drives a strongly first order EWPT is decoupled from EWB as the source of the BAU. As such, in what follows we will assume that the RH stop be at the TeV scale, and that a physical process that does not interfere with EWB drives a strongly first order EWPT. This assumption will allow us to explore wider regions of the parameter space relevant to EWB, although we shall also show the regions that would correspond to an LSP lighter than 125 GeV (dashed blue lines in fig.~\ref{fig:oh2} and \ref{fig:phim1}). From now on, in this manuscript we take all sfermions, including the RH stop, to be heavy, $m_{\tilde f}\gsim 1$ TeV. 

\section{Baryogenesis and Neutralino Relic (Over-)Abundance}\label{sec:oh2}

If the entropy density of the Universe is not a constant after the
freezeout of the net baryon density, the resulting baryon asymmetry
today will in general be affected. While episodes such as very
low-scale inflation (as low as the electroweak scale) have been
envisioned \cite{Lyth:1995ka, Knox:1992iy, Nardini:2007me}, this is
not the standard paradigm. However, an entropy injection episode
might actually be needed in a successful theory of supersymmetric
electroweak baryogenesis. We provide here an example and study the
implications for the underlying supersymmetric theory.

In the framework we consider here, where $R$-parity is conserved,
the LSP is stable. Given null
results in searches for a stable massive charged or
strongly-interacting particle \footnote{In particular, the number
density of a metastable negatively charged electroweak-scale
particle $\chi^-$ relative to entropy is constrained to be smaller
than $n_{\chi^-}/s \leq 3 \times 10^{-17}$, since it leads to
catalytic enhancement of the $^6{\rm Li}$ production.
\cite{Pospelov:2006sc}}, the LSP must be neutral, and its density
today must be compatible with (namely, at least not exceed) the
universal matter density to avoid over-closing the Universe. As
such, a relic, neutral LSP $\chi$ must have abundance $\Omega_\chi
h^2$ (in units of the critical density) less or equal to the dark
matter density $\Omega_{\rm DM}$. If this is not the case, either
the underlying supersymmetric theory is ruled out or a mechanism
must exist that dilutes away the excessive $\chi$ relic abundance.

Weakly interacting massive particles, such as the lightest
neutralino in MSSM scenarios where it is the LSP, undergo a rather
universal (i.e. independent of the details of masses and
interactions) freezeout process from the thermal bath in the early
universe that occurs around temperatures $T_{\rm freeze-out}\sim
m_\chi/(20\div25)$. In models of successful electroweak
baryogenesis, where the LSP cannot be heavier than $\sim 1$ TeV
\cite{Cirigliano:2006dg}, the freezeout temperature is thus always
below $50$ GeV. Therefore, freeze-out occurs typically at much lower
temperatures than those relevant for the electroweak phase
transition. In turn, this implies that if an entropy injection
episode is responsible for the dilution of an excessive thermal
relic density of neutralinos, the same entropy injection episode
will also dilute away the baryon number density produced at the
electroweak phase transition.

We are therefore left with two possibilities: either models with an
excessive neutralino thermal relic abundance are ruled out, or they
are salvaged by a modified cosmological expansion that inflates them
away \footnote{Interestingly, such a mechanism can be built in the
electroweak phase transition itself, see e.g. the recent analysis of Ref.~\cite{maxpaper}}.
However, if this second possibility is realized, the net baryon
asymmetry will also be diluted away, requiring a larger asymmetry to
be produced at the electroweak phase transition. It is this latter
possibility that we explore in this section.

%----------------------------------------------------------------
\begin{figure}[!h]
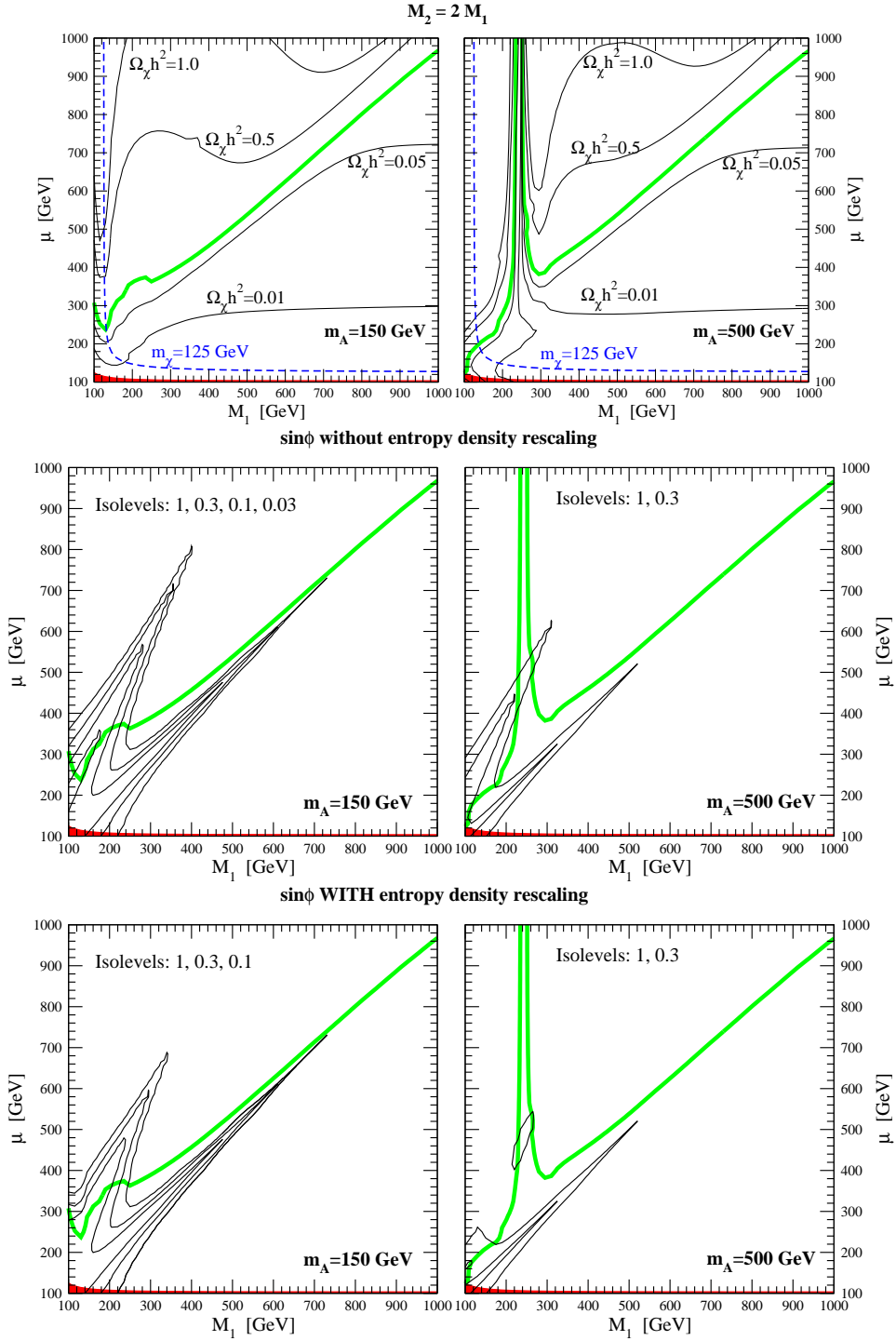

\begin{center}
%\hspace*{-1.cm}
\epsfig{file=plots/oh2_new.eps,height=6.05cm}
%\hspace*{-1.cm}
\epsfig{file=plots/yb_no_resc.eps,height=6.5cm}
%\hspace*{-1.cm}
\epsfig{file=plots/yb_resc.eps,height=6.5cm}
\end{center}
\caption{\it\small Iso-level curves for the lightest neutralino relic abundance (upper panels) and for the gaugino-higgsino CPV phase producing the central value of the BAU, without (central panels) and with (lower panels) entropy rescaling for models with over-abundant relic neutralinos. All panels refer to the ($M_1,\mu$) plane, with $m_A=150$ GeV in the left panels, and $m_A=500$ GeV for those to the right. The red shaded region is excluded by the non-observation of light neutralino pairs at LEP. The green bands correspond to a neutralino relic density consistent with the WMAP results. The blue dashed lines in the upper panels indicate a lightest neutralino mass of 125 GeV.}
\label{fig:oh2}
\end{figure}
%----------------------------------------------------------------
\clearpage

To illustrate the effect of enforcing a cosmologically acceptable
density of relic neutralinos, we consider in the upper two panels of
Fig.~\ref{fig:oh2} the thermal relic density of neutralinos in the
$(M_1,\mu)$ plane, for a scenario with heavy sfermions,
$\tan\beta=10$, $M_2=2\times M_1=M_3/3$ and $m_A=150$ GeV (top left panel)
and 500 GeV (top right panel). The green line corresponds to a
thermal relic abundance in accord with the inferred density of
cosmological dark matter. In the lower right corners of these two
plots, neutralinos only constitute a fraction of the cosmological
dark matter (unless some non-thermal production mechanism is
operative). Although this would mean one needs to postulate an
additional particle to provide the rest of the dark matter density,
an under-abundance of thermal neutralinos is cosmologically viable,
and there is no need to rescale the baryon asymmetry that gets
produced in those regions of the parameter space. The blue dashed lines in the upper panels indicate a lightest neutralino mass of 125 GeV.

In the lower right corner $\mu\lesssim M_1$, and the lightest
neutralino is higgsino-like, and it efficiently pair-annihilates
into $W^+W^-$ ad $ZZ$ pairs. Further contribution to the efficient
pair-annihilation processes in the early Universe also stems from
the co-annihilation of the quasi-degenerate next-to-lightest
higgsino-like neutralino and chargino. In the upper right portion of
the plots, the increasing bino-like content of the lightest
neutralino mass eigenstate suppresses annihilation to the SU(2)
gauge bosons, and only if a resonant annihilation channel occurs
(e.g. for $m_\chi\sim m_A/2$ in the left plot) can the relic
abundance be compatible with the upper limit to the cosmological
dark matter density. In that region, therefore, we will need to
rescale the neutralino density by at least a factor
$\Omega_\chi/\Omega_{\rm DM}$, the former factor being the thermal
neutralino relic density in units of the critical density.

Without accounting for such rescaling, the middle two panels in Fig.
1 show the values of $\sin\phi$ one would need to produce the
observed baryon asymmetry of the Universe. Each contour of the
double funnel-like region corresponds to a given value of ${\rm
sin}\phi$, where we take $\phi_1=\phi_2\equiv \phi$ under the assumption of phase universality. Going from the
outermost funnel to the innermost, successive contours correspond to
the values of $\sin\phi$ indicated in the panels \lq\lq isolevels").
The behavior of the contours is consistent with the results given in
Fig. 1 of our earlier work \cite{Cirigliano:2006dg}. Increasing the value
of $m_A$ suppresses the produced net baryon asymmetry, and therefore
the viable regions (those inside the funnels corresponding to the
resonant regions where $\mu\sim M_1$ or $\sim M_2$) shrink.

Accounting for the neutralino number density dilution and
transferring the dilution to the baryon number asymmetry as well
leads to a requirement of larger ${\rm sin}\phi$ (i.e. a larger
initial net baryon number density) in regions where
$\Omega_\chi\gtrsim\Omega_{\rm DM}$. This feature is illustrated in
the bottom two panels of Fig. 1. There, we find a suppression of the
$M_2\sim\mu$ funnel, where the lightest neutralino is dominantly
bino-like. For large values of $m_A$ (lower left panel), said funnel
essentially disappears, leaving bino-driven electroweak baryogenesis
\cite{Li:2008ez} as the only available option.

In the following section we further explore the phenomenological
consequences of entertaining a viable supersymmetric model from the
standpoint of cosmology, and we derive lower limits on the electron
and neutron EDM in the present scenario.

\section{A Lower Limit on the Electron and Neutron EDM}\label{sec:edm}

%----------------------------------------------------------------
\begin{figure}[!h]
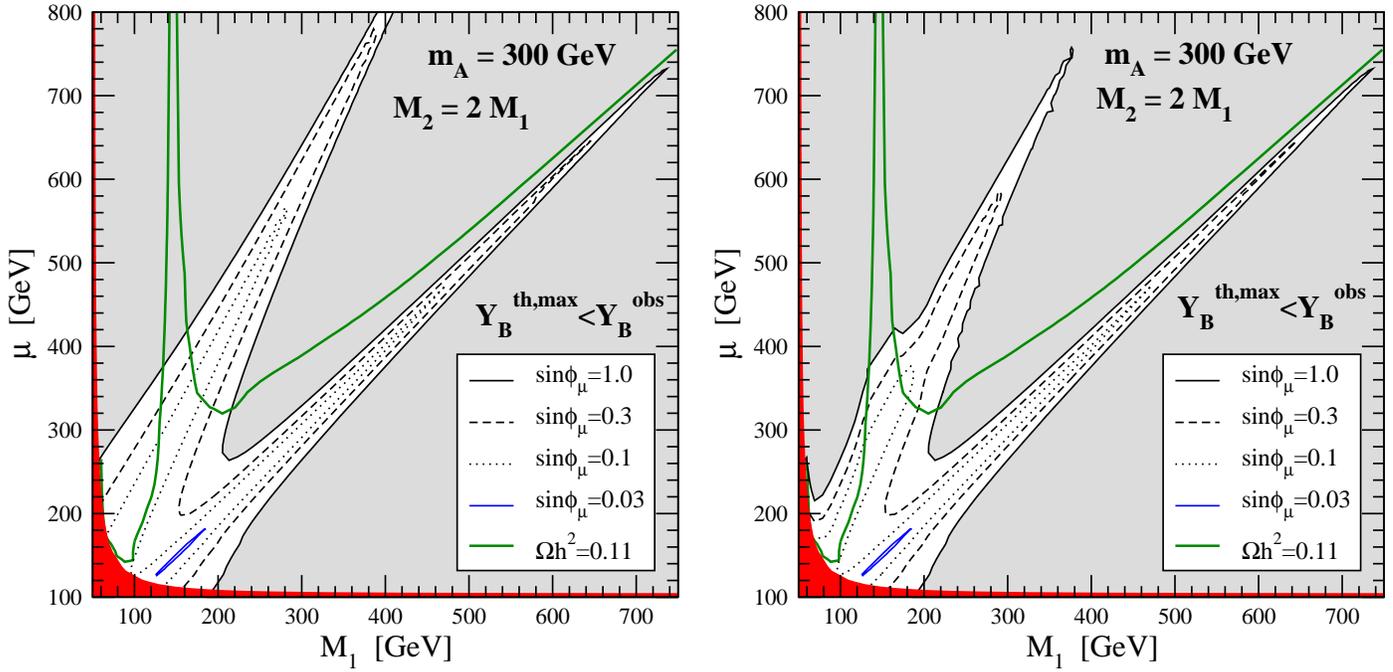

\begin{center}
\hspace*{-1.cm}\mbox{\epsfig{file=plots/300_phimu.eps,width=9.cm}\quad\epsfig{file=plots/300_phimuresc.eps,width=9.cm}}
\end{center}
\caption{\it\small Curves at constant CPV gaugino-higgsino phase, on the ($M_1,\mu$) plane, at $m_A=300$ GeV and $M_2=2M_1$, with (right) and without (left) entropy rescaling for overabundance relic neutralinos models. The grey region does not produce a large enough BAU through resonant processes, the red region is excluded by LEP searches for the lightest chargino, and the green lines indicate a neutralino relic abundance equal to the cold dark matter abundance in a standard cosmological setup.}
\label{fig:phimu300}
\end{figure}
%----------------------------------------------------------------

Fig.~\ref{fig:phimu300}, left, reiterates what is shown in the Fig.~\ref{fig:oh2}, but this time for a heavy Higgs sector set to $m_A=300$ GeV. As in the previous plots, we again assume here $M_2=2M_1$. The different shape in the contours depend on the assumed values for $m_A$. 
Again, the green lines indicate a thermal neutralino relic density in accord with the inferred cosmological dark matter density. The funnel where the lightest neutralino resonantly annihilates through the s-channel on-shell exchange of one of the heavy neutral Higgses appears near $M_1\sim m_A/2=150$ GeV. The left panel shows contours of constant values for the gaugino-higgsino phase needed to have the right amount of baryon asymmetry, neglecting the overabundance of neutralinos in the upper-left part of the parameter space. As indicated by the right panel, the effect of the required entropy dilution shrinks the available regions for those values of ($M_1,\mu$) where $\Omega_\chi\gtrsim\Omega_{\rm DM}$.

%----------------------------------------------------------------
\begin{figure}[!h]
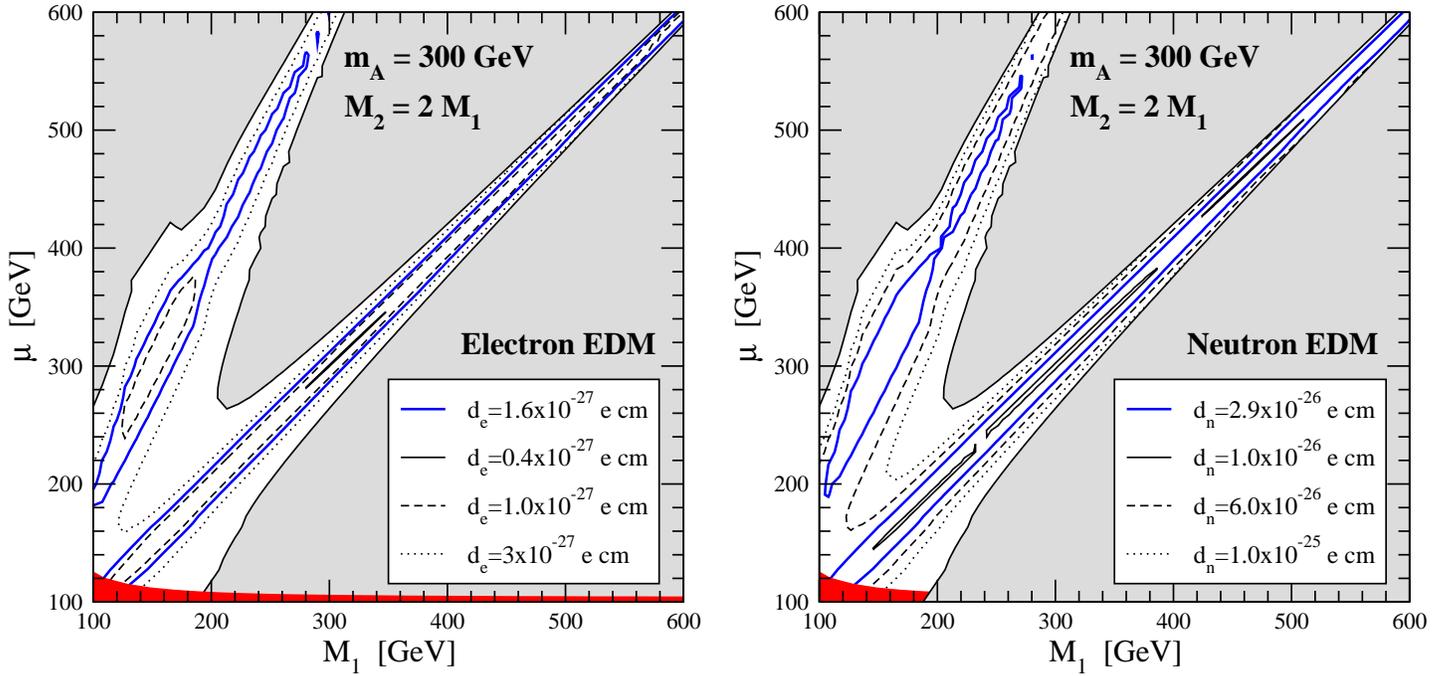

\begin{center}
\hspace*{-1.cm}\mbox{\epsfig{file=plots/300_de_new.eps,height=9.cm}\quad\epsfig{file=plots/300_dn.eps,height=9.cm}}
\end{center}
\caption{\it\small Curves of constant values for the electron (left) and for the neutron (right) electric dipole moment, on the same plane as Fig.~\ref{fig:phimu300}, right. The regions outside the blue contours are excluded by current experimental limits.}
\label{fig:edm300}
\end{figure}
%----------------------------------------------------------------

Focusing now on the physically motivated case of a re-scaled net
baryon asymmetry, Fig.~\ref{fig:edm300} shows lines of constant
values for the electron (left) and for the neutron (right) EDMs. As
one would expect, the largest values occur where the CP violating
phases are largest, i.e. near the outmost boundaries of the funnel
regions, while smaller values are obtained in the central part of
the funnels: there, the resonant source term allows for a lower CP
violating phase, resulting in smaller electric dipole moments. The only viable regions are therefore those {\em inside} the blue contours. In
summary, we find for the slice of parameter space we consider that
there is a lower limit to the electric dipole moment of the electron
around $0.4\times 10^{-27}$ e cm, roughly a factor 4 below the
current experimental limit; the analogous limit for the neutron EDM
sits at $10^{-26}$ e cm, i.e. only a factor 3 below current
sensitivity. We emphasize that these lower limits pertain only to
scenarios involving universality of phases: ${\rm Arg}(\mu M_1
b^*)={\rm Arg}(\mu M_2 b^*)\equiv\phi_\mu$. In section 5, we will relax
this unnecessary assumption.

%----------------------------------------------------------------
\begin{figure}[!h]
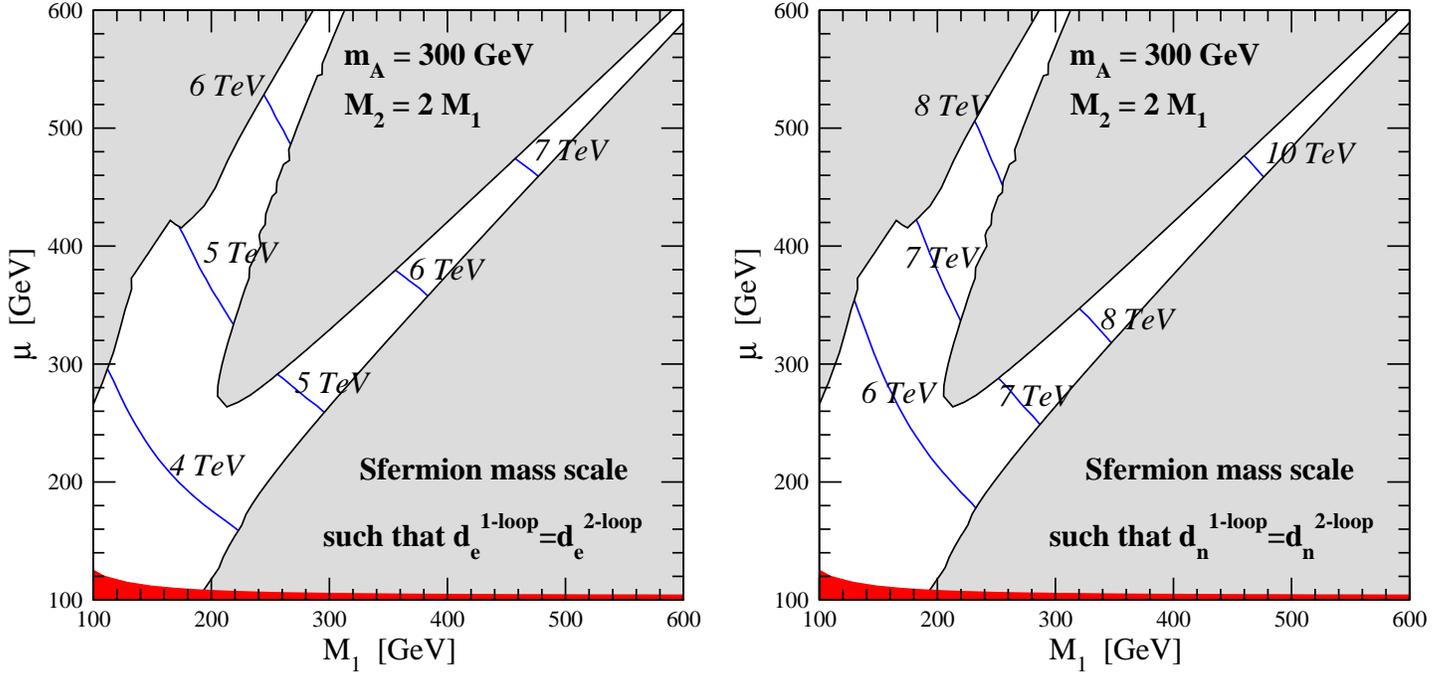

\begin{center}
\hspace*{-1.cm}\mbox{\epsfig{file=plots/300_sf_de.eps,height=9.cm}\quad\epsfig{file=plots/300_sf_dn.eps,height=9.cm}}
\end{center}
\caption{\it\small Curves indicating the values of a common sfermion mass scale such that one and two loop contributions to the electron (left) and to the neutron (right) EDM are equal, on the same plane as Fig.~\ref{fig:phimu300}, right. At a given sfermion mass scale indicated on the lines, points inside the funnel region lying closer to the origin correspond to  two-loop contributions that are larger in magnitude than one-loop contributions.  The converse holds for points lying closer to the funnel tips.}
\label{fig:edm_sf300}
\end{figure}
%----------------------------------------------------------------

Up to this point, we have considered only the (complete set
\cite{Li:2008kz}) two-loop contributions to the EDMs, since one-loop
contributions are negligible for sufficiently heavy 1st generation
sfermion masses. The following Fig. \ref{fig:edm_sf300} addresses
the question of how small the sfermion mass scale must be for one
loop contributions to become dominant over the two-loop ones. The
left and right panels of Fig. \ref{fig:edm_sf300} focus on the case of the EDM of the
electron and of the neutron, respectively. We set all CP violating phases to zero with the sole exception of the relative higgsino-gaugino phase, as in the rest of this study. The sfermion scale we refer to here alludes
naturally to the relevant sfermions, i.e. first generation sleptons
and squarks. Roughly, the size of the CP violating phase factors out from both the one- and two-loop contributions. The larger the values of $(M_1,\mu)$, the more suppressed the two-loop contributions: as a result, the sfermion mass scale that would produce a comparable EDM can also be larger, as the iso-level contour labels show. Fixing the sfermion mass scale to e.g. 6 TeV, points to the upper-right of the lines have a larger one-loop contribution than the two-loop one, while points to the lower-left side of the lines have  a larger two-loop contribution.

In summary, we find that the one-loop contributions to the electron
EDM start to be significant when the mass scale is around 4-7 TeV,
while squarks as heavy as 10 TeV can give significant contributions
at the one-loop level to the neutron EDM. This implies, in
particular, that the neutron EDM is overall more sensitive to
one-loop contributions than the electron EDM

\section{EDM in the bino-driven electroweak baryogenesis scenario}\label{sec:bino}

As pointed out by some of us in Ref.~\cite{Li:2008ez}, it is possible for MSSM baryogenesis
to remain viable even if the next generation of EDM searches yield null results. Theoretically,
this scenario requires that one relaxes the
assumption of universality between the relative bino-higgsino and
wino-higgsino phase, i.e. $\phi_1 \ne \phi_2$ and  work in the
limit of heavy 1st generation sfermion masses. In this way, as long
as $\phi_2$ is sufficiently small, EDMs are suppressed, while
bino-driven electroweak baryogenesis, efficient if the resonant
condition $\mu\sim M_1$ is approximately valid, can proceed with
large bino-higgsino phases $\phi_1$. The latter, as shown and
explained in detail in Ref.~ \cite{Li:2008ez}, does not lead to
excessively large contributions to the EDMs, even if $|\sin\phi_1|$ is of order
one.

%----------------------------------------------------------------
\begin{figure}[!h]
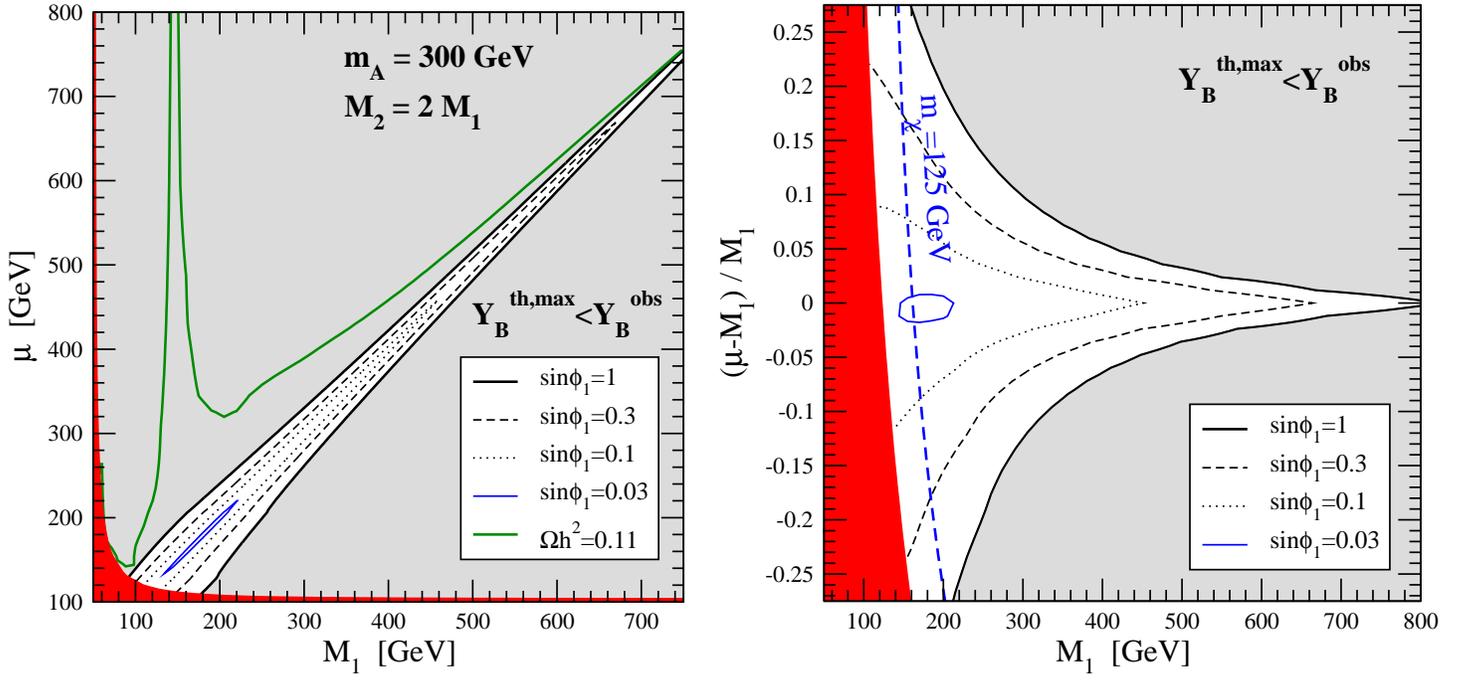

\begin{center}
\hspace*{-1.cm}\mbox{\epsfig{file=plots/300_phim1_bau.eps,height=9.cm}\quad\epsfig{file=plots/300_phim1_deltamu_new.eps,height=9.cm}}
\end{center}
\caption{\it\small Curves indicating constant values of the relative bino-higgsino phase $\phi_1$ that produces the right BAU through resonant processes, on the same plane as fig.~\ref{fig:phimu300} (left), and on the plane defined by $M_1$ and by the relative bino-higgsino mass splitting $(\mu-M_1)/M_1$. The grey region does not produce a large enough BAU through resonant processes, the red region is excluded by LEP searches for the lightest chargino. In the left panel, the green lines indicate a neutralino relic abundance equal to the cold dark matter abundance in a standard cosmological setup. In the right panel, the blue dashed line indicates a lightest neutralino mass of 125 GeV.}
\label{fig:phim1}
\end{figure}
%----------------------------------------------------------------

In the bino-driven portion of the parameter space,
$\Omega_\chi\lesssim\Omega_{\rm DM}$, and no entropy dilution is
needed to wash out excess relic neutralinos. We illustrate this
point in the left panel of Fig. 5, where we assume $\phi_2=0$ and
show the values of $\phi_1$ for which we generate the observed
baryon asymmetry. Since resonant bino-driven EWB is most efficient
in the region for $\mu \sim M_1$, it is useful to re-cast our results in terms
of $M_1$ and the relative mass-splitting $(\mu-M_1)/M_1$. In the
right panel of Fig. 5, we  show the bino-driven 
region (i.e. the region where $\mu\sim M_1$) in the $(M_1, (\mu-M_1)/M_1)$ plane. As the contours for
successively smaller values of ${\rm sin}\phi$ indicate, a lighter
overall mass scale and smaller relative mass splitting are needed to
reproduce the observed $Y_B$ as ${\rm sin}\phi$ is decreased.

%----------------------------------------------------------------
\begin{figure}[!h]
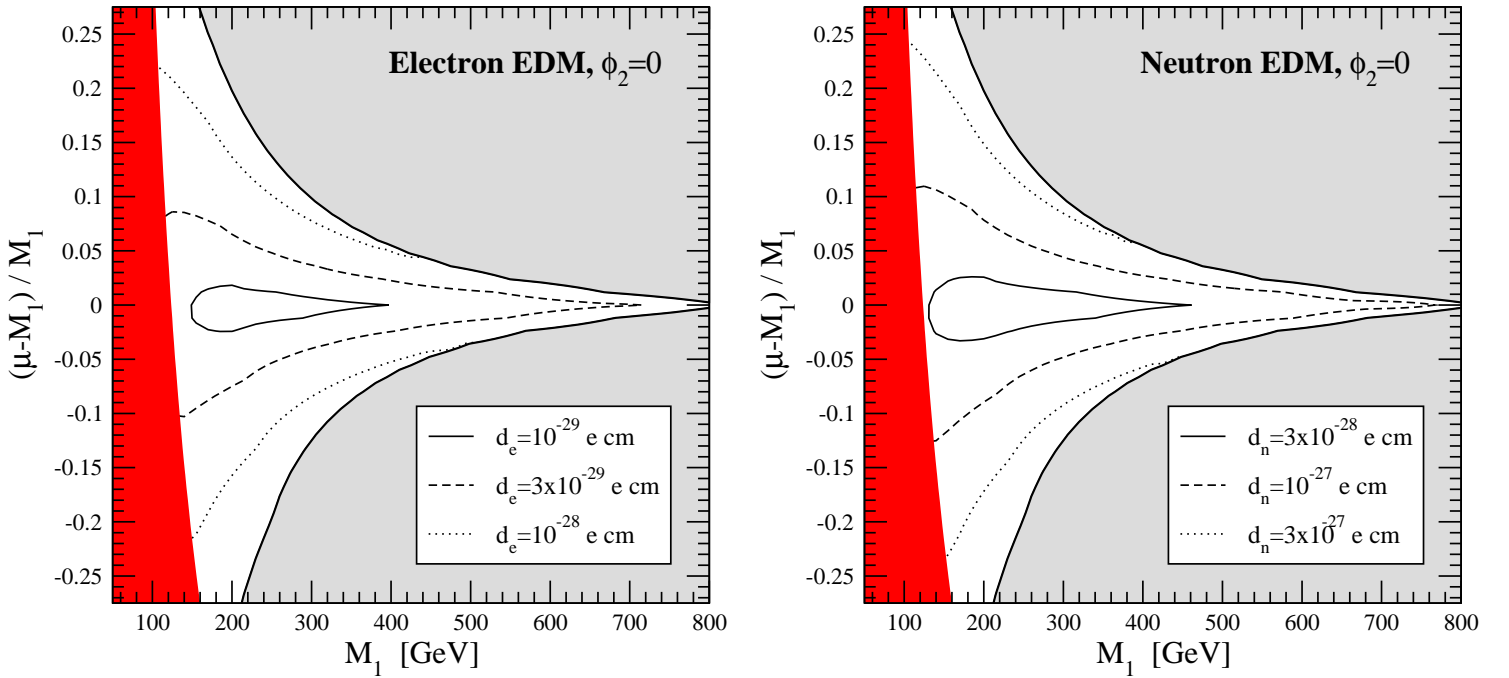

\begin{center}
\hspace*{-1.cm}\mbox{\epsfig{file=plots/300_phim1_deltamu_de.eps,height=9.cm}\quad\epsfig{file=plots/300_phim1_deltamu_dn.eps,height=9.cm}}
\end{center}
\caption{\it\small Curves indicating constant values for the electron (left) and for the neutron (right) electric dipole moment, on the same plane as fig.~\ref{fig:phim1}, right. The CPV phase $\phi_1$ is set to the value giving the right BAU, while $\phi_2$ (the relative wino-higgsino phase) is set to 0.}
\label{fig:edm_phim1}
\end{figure}
%----------------------------------------------------------------

We employ the same parameter space (the $(M_1,(\mu-M_1)/M_1)$ plane)
to illustrate that even in the case of bino-driven EWB we obtain an
absolute lower limit on the size of the electron and the neutron EDM
that would be consistent with this scenario. Here we give contours
of constant $d_e$ (left panel) and $d_n$ (right panel), rather than
constant ${\rm sin}\phi_1$, consistent with $Y_B$. The contours in
the left panel of Fig.~\ref{fig:edm_phim1}, left, indicate that the
minimal electron EDM is around $10^{-29}$ e cm, roughly a
factor 200 below present limits. The projected sensitivity
improvement to cover this slice of the bino-driven electroweak
baryogenesis scenario with neutron EDM searches is instead of around
a factor 100, as shown in the right panel.

\section{The $(\tan\beta,m_A)$ Plane}\label{sec:tbma}

Thus far, our exploration of the supersymmetric parameter space
compatible with EWB has focused largely on the parameter space
defined by the masses of those particles directly entering the
source terms in the production of the baryon asymmetry  at the
electroweak phase transition. We now turn our attention to two other
parameters that play a crucial role in the dynamics of baryon number
generation at electroweak temperatures: the ratio of the vacuum expectation values 
of the neutral componets of the  two SU(2) Higgs doublets, $\tan\beta$, and
the mass of the CP-odd Higgs, $m_A$. The former enters in the
calculation of $Y_B$ via not only directly in the sources via the
various particle couplings, including Yukawa couplings, but also via
the effects described in Ref.~\cite{Chung:2008aya,Chung:2009qs}. In
the latter work, it was observed that for moderate values of ${\rm
tan}\beta$, the timescale for b (squark) and tau (s)lepton Yukawa
interactions can be shorter than the timescale associated with
diffusion ahead at the bubble wall. Consequently, for sufficiently
light right-handed sbottoms and staus, the dynamics of these
sfermions can substantially alter the transport dynamics that govern
the conversion of Higgs-Higgsino CP-violating asymmetries into the
(s)quark sector asymmetries. The mass parameter $m_A$ also enters in
the calculation of $Y_B$ through the parameter $\Delta\beta$, to
which $Y_B$ is linearly proportional. Here, we treat the dependence
of $Y_B$ on $m_A$ according to the two-loop results presented in
\cite{Moreno:1998bq}, and the dependence on $\tan\beta$ according to
the full treatment outlined in \cite{Chung:2009qs} \footnote{Note
that we do not consider the effects of varying $m_A$ on the strength
of the electroweak phase transition since in the latest work of Ref.
\cite{Carena:2008rt}, the pseudoscalar A was integrated out of the
effective theory.}.

So far, the ($\tan \beta, m_A$)  parameter space for this problem has not  
been extensively explored. 
For illustrative purposes,  we adopt here the universal phase assumption  
and we show that there  can be strong bounds in this space that have not 
been emphasized in previous work. At  the end of this analysis, we 
discuss briefly our expectations for the case of non-universal phases.

%----------------------------------------------------------------
\begin{figure}[!h]
\begin{center}
\hspace*{-1.cm}\epsfig{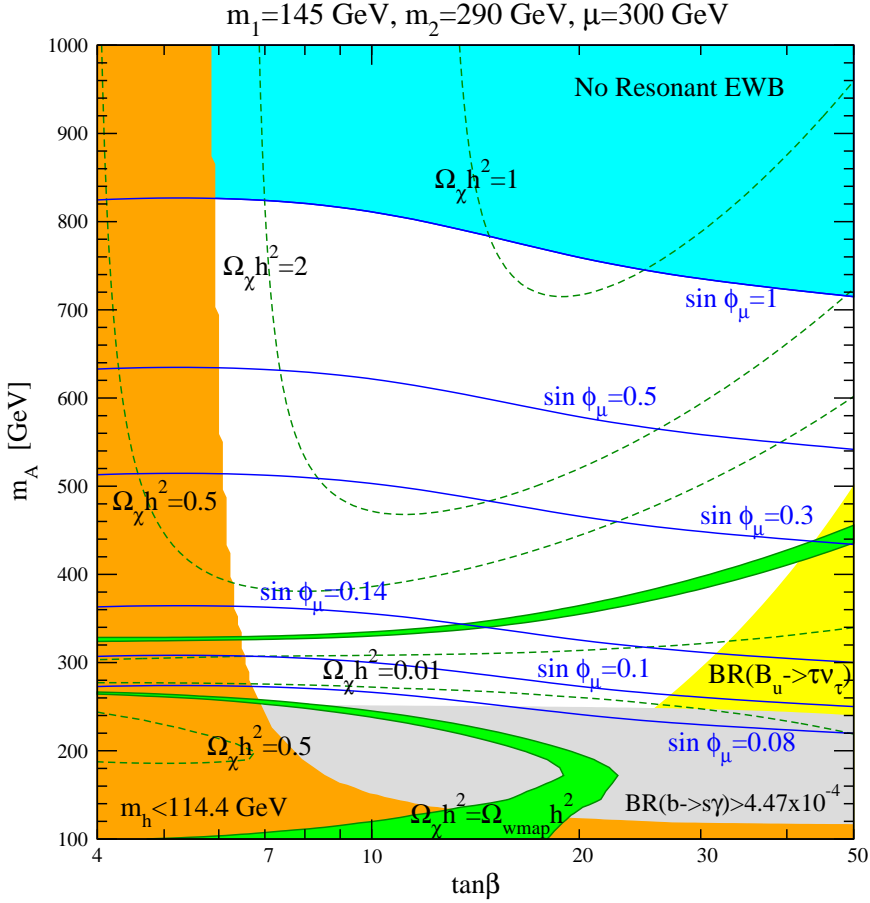}
\end{center}
\caption{\it\small The $(\tan\beta, m_A)$ plane: the orange region
is ruled out by LEP limits on the lightest Higgs, while the grey
region is ruled out by excessive supersymmetric contributions to the
inclusive $b\to s\gamma$ branching ratio and the yellow region is ruled out by bounds on the BR($B_u\to\tau\nu_\tau$). In the blue region at the
top the generated BAU is too small, while the green contours
indicate the 2-$\sigma$ preferred range for relic neutralinos to be
the dark matter. We also show contours of constant values for the
neutralino relic abundance (green dashed lines) and for the value of
the gaugino-higgsino CPV phase $\sin\phi_\mu$ (solid blue lines) such that
electroweak baryogenesis successfully accounts for the BAU.}
\label{fig:tbma}
\end{figure}
%----------------------------------------------------------------

In Fig. 7, we show the constraints on  the $({\rm tan}\beta,m_A)$
parameter space from various phenomenological and cosmological
considerations. We set the gaugino and higgsino mass parameters to
$M_1=145$ GeV, $M_2=290$ GeV and $\mu=300$ GeV. 
As discussed above, we also assume a
common gaugino-higgsino phase $\phi = {\rm Arg}(\mu M_j b^*)$, $j=1,2$.
In the orange region to the left and bottom of the plane, the Higgs
mass is lower than the limit set by LEP-II of 114.4 GeV (this limit applies for the heavy sfermion sector we consider in the present analysis). In
addition, in the grey area the supersymmetric contribution to the
$b \to s \gamma$ decay 
exceeds the experimentally allowed value. 

At large values of $\tan\beta$, a significant constraint stems from MSSM contributions to the $B_u\to\tau\nu_\tau$ decay mode \cite{bdecay1,bdecay2}. This branching ratio, in turn, depends on SUSY-QCD corrections to the charged Higgs boson coupling to fermions. These corrections are asymptotically small in the limit of heavy sfermions, and, for our choices of parameters, the relevant quantity $\epsilon_0$ in Eq.~(6) of Ref.~\cite{bdecay1} is always smaller than 0.01. Given that in the present analysis we mostly focus on the limit of heavy sfermions to avoid one loop contributions to the EDMs, we set $\epsilon_0=0$, which entails slightly more stringent constraints than for a small, finite positive value. Imposing the 2-$\sigma$ limit to the ratio of the MSSM to SM contribution $0.53<R_{\tau\nu_\tau}<2.03$ \cite{bdecay1}, two disconnected regions of the ($\tan\beta,m_A$) parameter space are ruled out. As shown in Ref.~\cite{bdecay2}, the portion of parameter space between these disconnected regions is actually also ruled out by constraints on MSSM contributions to the ratio of $K\to\mu\nu_\mu$ and $\pi\to\mu \nu_\mu$ branching ratios. For the range of $\tan\beta$ shown in Figs. ~7-10, the latter region also lies within the $b\to s\gamma$ exclusion region. We shade in yellow the remaining excluded parameter space implied by the $B_u\to\tau\nu_\tau$ constraints. 

The green dashed lines correspond to various, constant values of the lightest neutralino
relic abundance $\Omega_\chi h^2$ (as indicated by the labels).
Within the thick green bands, the thermal neutralino relic abundance
is within the 2$\sigma$ WMAP range for the average cold dark matter
density. Finally, the blue lines indicate which value of the phase
$\phi$ is needed for the particular point in parameter space to
produce the observed $Y_B$. For these lines, we {\em do not} include
the dilution of over-abundance neutralinos. In the upper region of
the plot, shaded in cyan, no resonant electroweak baryogenesis can
produce enough baryon asymmetry, even neglecting the effect of
entropy dilution: this therefore sets, for the particular values of
gaugino and higgsino parameters we picked here, an upper limit of
$m_A \leq 800$ GeV.

%----------------------------------------------------------------
\begin{figure}[!h]
\begin{center}
\hspace*{-1.cm}\epsfig{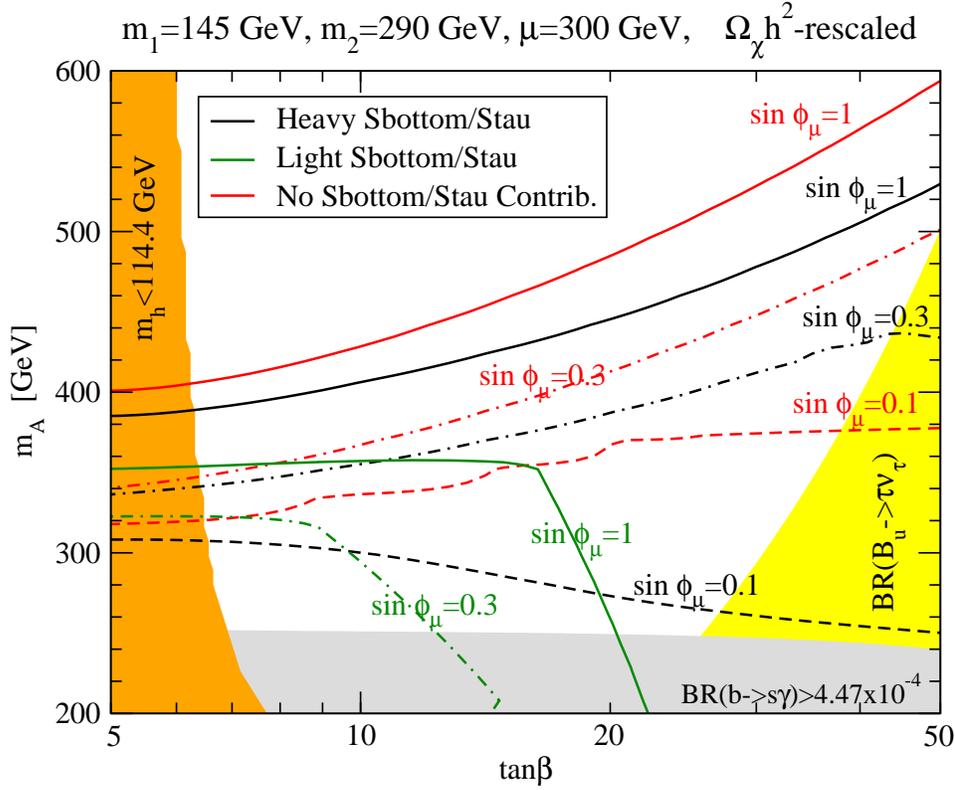}
\end{center}
\caption{\it\small The same plane as in fig.~\ref{fig:tbma}, but
taking into account the effect of relic neutralino dilution in
models over-producing dark matter on the CPV phases needed to
achieve successful electroweak baryogenesis. The color code is the
same as in fig.~\ref{fig:tbma}. We indicate with different colors
lines corresponding to different assumptions on the mass of the
Sbottom and of the Stau sfermions. Namely, our default assumption of
heavy sfermions is indicated in black, while light (300 GeV)
sfermions correspond to the green line. Neglecting the effect of the
Yukawa interactions involving sbottoms and staus produces the lines
indicated in red.} \label{fig:tbma_resc}
\end{figure}
%----------------------------------------------------------------

In Fig.~\ref{fig:tbma}, as in the remainder of this paper unless
otherwise specified, we have assumed heavy squarks and we have
neglected the effect of down-type sfermions (effectively assuming
heavy sbottoms and staus). We study in fig.~\ref{fig:tbma_resc} the
impact of two effects: (i) rescaling of the net baryon asymmetry for
parameter space points with a relic thermal neutralino abundance in
excess of the dark matter density, and (ii) factoring in finite
masses for third generation sfermions. 
The lines in  Fig.~\ref{fig:tbma_resc} represent iso-level contours of $\phi$ 
(corresponding to the observed $Y_B$) under different assumptions. 
The red lines 
correspond to the limit of super-heavy down-type
sfermions, after rescaling for over-dense neutralino relic
densities. As evident, the parameter space shrinks with respect to 
the case without density rescalling (Fig~\ref{fig:tbma}), and for
reasonably low values of $\tan\beta$, the maximal value $m_A$ can
take is below 600 GeV. Note that in contrast to the isolevel contours of Fig.~\ref{fig:tbma}, those of Fig.~\ref{fig:tbma_resc} increase  monotonically with $\tan\beta$. 

The black lines give the isolevel curves when the effect of bottom and top Yukawa interactions are included in
the transport dynamics,  for right-handed sbottom and stau masses
equal to 1 TeV.  Inclusion of these RH sfermions suppresses the baryon asymmetry, implying a need for a lighter $m_A$ for a given value of $\sin\phi_\mu$. The impact of the RH third generation sfermions is even more pronounced when they are relatively light. To illustrate, we show in green the isolevel contours for 300 GeV ${\tilde b}_R$ and ${\tilde\tau}_R$ masses.  The appearance of \lq\lq elbows" in these contours arises because the suppression of $Y_B$ in the presence of light ${\tilde b}_R$ and ${\tilde\tau}_R$ grows with $\tan\beta$.  For the relatively flat portion of the green curves at lower $\tan\beta$, the suppression due to the ${\tilde b}_R$ and ${\tilde\tau}_R$ essentially compensates for the monotonic increase that would otherwise occur when these sfermions are heavy (the black curves). Eventually, the light RH sfermion suppression takes over, pushing the value of $Y_B$ below the experimental value and necessitating a smaller value of $m_A$ as needed to enhance the CPV source in the transport equations. This transition point corresponds to the elbow in a given isolevel curve.  Interestingly, the resulting parameter space allowed by the
requirement of successful EWB for 300 GeV sfermions is now {\em
bounded}, with $\tan\beta\lesssim20$ and $m_A\lesssim350$ GeV. Note
that we have not yet imposed EDM constraints.

%----------------------------------------------------------------
\begin{figure}[!h]
\begin{center}
\hspace*{-1.cm}\epsfig{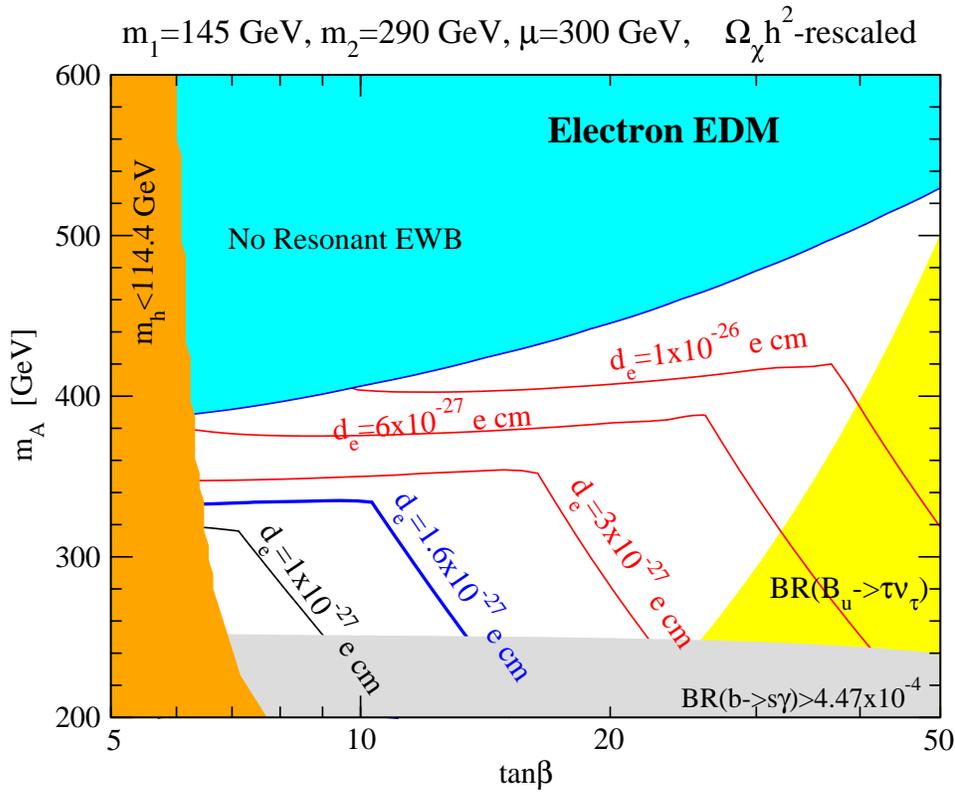}
\end{center}
\caption{\it\small Contours of constant values for the two-loop electron EDM, on the same plane as the one shown in fig.~\ref{fig:tbma_resc} (i.e including entropy rescaling). The blue line indicates the current experimental limit: the viable parameter space lies below that curve.}
\label{fig:tbma_de}
\end{figure}
%----------------------------------------------------------------

\begin{figure}[!t]
\begin{center}
\hspace*{-1.cm}\epsfig{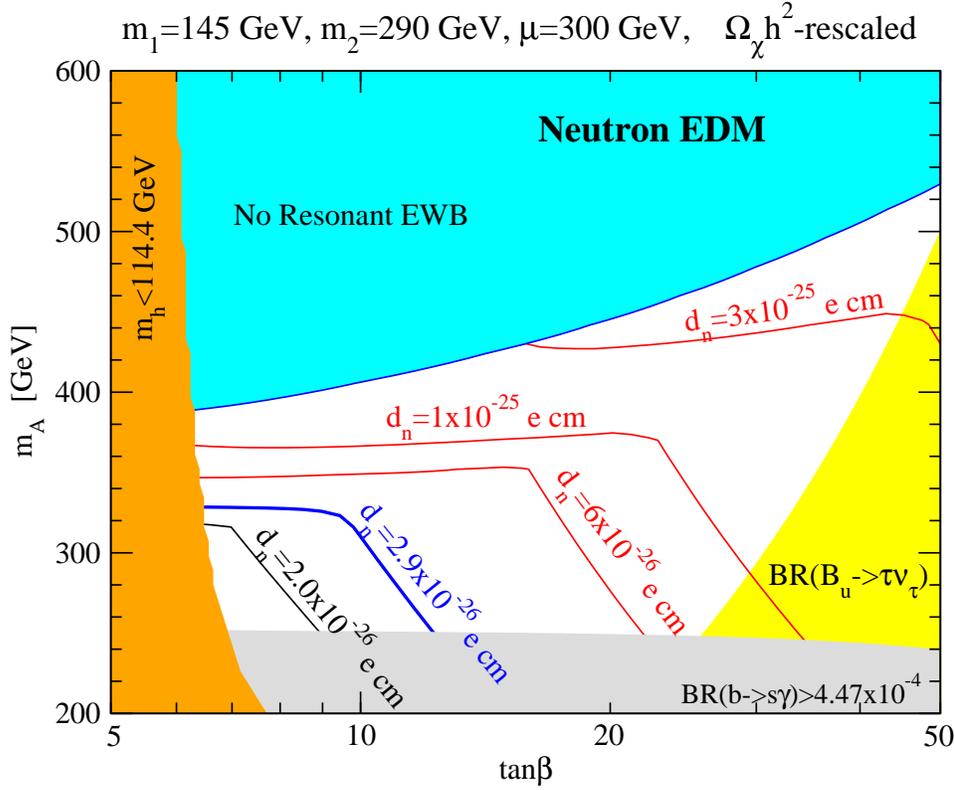}
\end{center}
\caption{\it\small As in fig.~\ref{fig:tbma_de}, this time for the two-loop neutron electric dipole moment.}
\label{fig:tbma_dn}
\end{figure}

We now investigate the size of the permanent EDMs of the electron (Fig~\ref{fig:tbma_de}) and of the neutron (Fig~\ref{fig:tbma_dn}) over the same parameter space, taking into account the relic density rescaling, but still assuming super-heavy third generation sfermions. Our results are thus conservative here, given that including the effect of lighter sfermions would suppress $Y_B$ and force larger CP violating phases.

The shape of the contours in the figures illustrates how the size of the two-loop EDMs depends on both the size of $m_A$ (larger values suppress the two loop contributions where the heavy Higgs sector enters) and on $\tan\beta$.
In addition, the EDMs also linearly depend on the size of the CP violating phases (that partly compensate for the suppression induced by larger values of $m_A$). As a result, the parameter space is bounded, and there is an absolute lower limit to the size of the EDMs on this slice of parameter space as well. Specifically, for the choice of parameters we make here, the electron EDM exceeds the current experimental limit for $m_A\gtrsim 330$ GeV and $\tan\beta\gtrsim15$, with even more stringent constraints from the neutron EDM. The smallest the electron EDM can get on this slice of parameter space is 
a few $\times 10^{-28}$ e cm, less than one order of magnitude below the current experimental sensitivity. The neutron EDM (fig.~\ref{fig:tbma_dn})
 can at most be a factor of a few below the current sensitivity.

In short, the study of the $(\tan\beta,m_A)$  parameter space enforces the conclusion we obtained from the study of the orthogonal gaugino-higgsino parameter space: the size of the electron and of the neutron EDM, even only at the two-loop level, is bounded from below, and EWB will be thoroughly tested in the next generation of EDM search experiments.
For the case of non-universal phases, we expect the EDM constraints on 
the $(\tan \beta,m_A)$   plane  to be weaker, given the reduced  
sensitivity of EDMs to the bino phase (assuming it is the one  
responsible for resonant EWB). With an improvement of $O(100 )$ in EDM  
sensitivity, however, we expect the restrictions on the parameter  
space in this plane will be similar to those shown for the universal  
phase case.

%----------------------------------------------------------------
%----------------------------------------------------------------

\section{Discussion and Conclusions}\label{sec:conclusions}

In this paper we continued our program of investigating the
phenomenology of supersymmetric models with successful electroweak
baryogenesis. Our primary findings are:
\begin{enumerate}
\item Regions of parameter space with over abundant thermal relic neutralino production are either cosmologically ruled out or must occur in conjunction with a late-time reheating episode. In the latter instance, the consequent dilution of both the neutralino relic density and  baryon number asymmetry implies that the baryon number density at the EWPT must have been larger. These considerations lead to more demanding requirements on the MSSM parameters in the domain of resonant wino-Higgsino driven baryogenesis. As a result, we find that the parameter space for resonant  wino-Higgsino driven baryogenesis is significantly more constrained than previously realized. In contrast, the requirements for successful bino-Higgsino driven baryogenesis are largely unaffected. 
\item We explored for the first time the parameter space defined by the pair $m_A$ and $\tan\beta$. We find that a general upper limit exists to the maximal value the heavy Higgs mass scale can take, and this limit is even stronger if relic density dilution effects are taken into account. If third generation sfermions have a low mass scale, then an upper limit to $\tan\beta$ also emerges due role played by the third generation (s)fermions in the transport dynamics. 
\item General lower limits exist on both the electron and the neutron EDMs, which are even stronger when the dilution effects are taken into account. Assuming phase universality, these lower bounds lie below the current experimental bounds by less than one order of magnitude. The non-observation of electron and neutron EDMs in experiments having ten times better sensitivity would likely rule out  MSSM baryogenesis under the assumption of phase universality. These prospects can be seen from both the accessible regions of the ($\mu$,$M_{1,2}$) parameter space and in the ($\tan\beta$,$M_A$) plane. 
\item If one relaxes the assumption of phase universality, then the expected lower bounds on $|d_e|$ and $|d_n|$ consistent with the scenario of resonant, bino-driven MSSM baryogenesis are roughly ten times smaller than we find under the assumption of phase universality. Because the strength of the CP-violating sources used in our analysis is likely to over-predict the strength of these sources that would likely result from  a fully consistent, all-orders Higgs vev resummation, we expect that the resonant bino-driven scenario will be testable with the next generation electron and neutron EDM searches that aim for two orders of magnitude improved sensitivity. 
\end{enumerate}
The general picture that emerges from our present phenomenological update
is one where MSSM electroweak baryogenesis is a
predictive framework that will be conclusively tested in the near
future. Indeed, the parameter space we combed in this analysis is  devoid of regions where the predictions for
 specifically EDMs are unobservably small. Thus, the planned  improvements in experimental sensitivity will put the
entire setup under thorough scrutiny. Although there remain important open theoretical questions pertaining to the transport dynamics, such as the $\tan\beta$-dependence of the bubble wall profiles and consequences of a consistent resummation of the background field for the CP-violating sources, we expect that our general conclusions regarding the testability of MSSM baryogenesis will continue to hold after these issues are conclusively addressed.

%******************************************************************************

\vspace*{1cm}
\noindent{{\bf Acknowledgments} } \\
\noindent 

\noindent We thank Sean Tulin for discussions and for providing help with part of the numerical results. The work of VC    is supported by the Nuclear Physics Office of the U.S.
Department of Energy under Contract No. DE-AC52-
06NA25396 and by the LDRD program at Los Alamos
National Laboratory.
MJRM and YL were supported in part by U.S. Department of Energy
contract DE-FG02-08ER41531 and by the Wisconsin Alumni Research
Foundation. MJRM also thanks the Aspen Center for Physics and TRIUMF
Theory Group where part of this work was completed.
S.P. is partly supported by an Outstanding Junior Investigator Award from the U.S. Department of Energy, 
Office of Science, High Energy Physics, DoE Contract DEFG02-04ER41268, and by NSF Grant PHY-0757911.

%******************************************************************************
%******************************************************************************


\begin{thebibliography}{200}
\small

%%%%%%%%%%%%%%%%%%%%%%%%%%%%
%
%   General Refs
%
%%%%%%%%%%%%%%%%%%%%%%%%%%%%


\bibitem{Yao:2006px}
  W.~M.~Yao {\it et al.}  [Particle Data Group],
  %``Review of particle physics,''
  J.\ Phys.\ G {\bf 33} (2006) 1.

\bibitem{Dunkley:2008ie}
  J.~Dunkley {\it et al.}  [WMAP Collaboration],
  %``Five-Year Wilkinson Microwave Anisotropy Probe (WMAP) Observations:
  %Likelihoods and Parameters from the WMAP data,''
  arXiv:0803.0586 [astro-ph].
  %%CITATION = ARXIV:0803.0586;%%

\bibitem{Sakharov:1967dj}
A.~D.~Sakharov,
%``Violation Of CP Invariance, C Asymmetry, And Baryon Asymmetry Of The
%Universe,''
Pisma Zh.\ Eksp.\ Teor.\ Fiz.\  {\bf 5}, 32 (1967)
[JETP Lett.\  {\bf 5}, 24 (1967)].

\bibitem{leph}
  R.~Barate {\it et al.}  [LEP Working Group for Higgs boson searches],
  %``Search for the standard model Higgs boson at LEP,''
  Phys.\ Lett.\ B {\bf 565} (2003) 61
  [arXiv:hep-ex/0306033].

\bibitem{noewb}
  G.~R.~Farrar and M.~E.~Shaposhnikov,
  %``Baryon asymmetry of the universe in the standard electroweak theory,''
  Phys.\ Rev.\ D {\bf 50} (1994) 774
  [arXiv:hep-ph/9305275].

\bibitem{Bochkarev:1987wf}
  A.~I.~Bochkarev and M.~E.~Shaposhnikov,
  %``Electroweak Production Of Baryon Asymmetry And Upper Bounds On The Higgs
  %And Top Masses,''
  Mod.\ Phys.\ Lett.\ A {\bf 2} (1987) 417.

\bibitem{baerbook}
See e.g.: Baer, H., \& Tata, X.\ 2006, Weak Scale Supersymmetry, by Howard Baer and Xerxes Tata, pp.~556.~Cambridge University Press, May 2006.~ISBN-10: 0521857864.~ISBN-13: 9780521857864,  


\bibitem{Griffith:2009zz}
  W.~C.~Griffith, M.~D.~Swallows, T.~H.~Loftus, M.~V.~Romalis, B.~R.~Heckel and E.~N.~Fortson,
  %``Improved Limit on the Permanent Electric Dipole Moment of Hg-199,''
  Phys.\ Rev.\ Lett.\  {\bf 102}, 101601 (2009).
  %%CITATION = PRLTA,102,101601;%%

\bibitem{Carena:2008rt}
  M.~Carena, G.~Nardini, M.~Quiros and C.~E.~M.~Wagner,
  %``The Effective Theory of the Light Stop Scenario,''
  JHEP {\bf 0810}, 062 (2008)
  [arXiv:0806.4297 [hep-ph]].
  %%CITATION = JHEPA,0810,062;%%

\bibitem{Li:2008kz}
  Y.~Li, S.~Profumo and M.~Ramsey-Musolf,
  %``Higgs-Higgsino-Gaugino Induced Two Loop Electric Dipole Moments,''
  Phys.\ Rev.\  D {\bf 78}, 075009 (2008)
  [arXiv:0806.2693 [hep-ph]].
  %%CITATION = PHRVA,D78,075009;%%

\bibitem{Li:2008ez}
  Y.~Li, S.~Profumo and M.~Ramsey-Musolf,
  %``Bino-driven Electroweak Baryogenesis with highly suppressed Electric Dipole
  %Moments,''
  Phys.\ Lett.\  B {\bf 673}, 95 (2009)
  [arXiv:0811.1987 [hep-ph]].
  %%CITATION = PHLTA,B673,95;%%

\bibitem{Chung:2009qs}
  D.~J.~H.~Chung, B.~Garbrecht, M.~J.~Ramsey-Musolf and S.~Tulin,
  %``Supergauge interactions and electroweak baryogenesis,''
  arXiv:0908.2187 [hep-ph], arXiv:0905.4509 [hep-ph].
  %%CITATION = ARXIV:0908.2187;%%

\bibitem{Carena:2008vj}
  M.~Carena, G.~Nardini, M.~Quiros and C.~E.~M.~Wagner,
  %``The Baryogenesis Window in the MSSM,''
  Nucl.\ Phys.\  B {\bf 812}, 243 (2009)
  [arXiv:0809.3760 [hep-ph]].
  %%CITATION = NUPHA,B812,243;%%


\bibitem{Lee:2004we}
  C.~Lee, V.~Cirigliano and M.~J.~Ramsey-Musolf,
  %``Resonant relaxation in electroweak baryogenesis,''
  Phys.\ Rev.\ D {\bf 71} (2005) 075010
  [arXiv:hep-ph/0412354].


\bibitem{Cirigliano:2006wh}
  V.~Cirigliano, M.~J.~Ramsey-Musolf, S.~Tulin and C.~Lee,
  %``Yukawa and tri-scalar processes in electroweak baryogenesis,''
  Phys.\ Rev.\  D {\bf 73}, 115009 (2006)
  [arXiv:hep-ph/0603058].
  %%CITATION = PHRVA,D73,115009;%%

\bibitem{Chung:2008aya}
  D.~J.~H.~Chung, B.~Garbrecht, M.~J.~Ramsey-Musolf and S.~Tulin,
  %``Yukawa Interactions and Supersymmetric Electroweak Baryogenesis,''
  Phys.\ Rev.\ Lett.\  {\bf 102}, 061301 (2009)
  [arXiv:0808.1144 [hep-ph]].
  %%CITATION = PRLTA,102,061301;%%

\bibitem{Carena:2002ss}
  M.~Carena, M.~Quiros, M.~Seco and C.~E.~M.~Wagner,
  %``Improved results in supersymmetric electroweak baryogenesis,''
  Nucl.\ Phys.\ B {\bf 650} (2003) 24
  [arXiv:hep-ph/0208043].
  %%CITATION = HEP-PH 0208043;%%

\bibitem{Konstandin:2004gy}
  T.~Konstandin, T.~Prokopec and M.~G.~Schmidt,
  %``Kinetic description of fermion flavor mixing and CP-violating sources  for
  %baryogenesis,''
  Nucl.\ Phys.\ B {\bf 716}, 373 (2005)
  [arXiv:hep-ph/0410135].
  %%CITATION = HEP-PH 0410135;%%

\bibitem{Konstandin:2003dx}
  T.~Konstandin, T.~Prokopec and M.~G.~Schmidt,
  %``Axial currents from CKM matrix CP violation and electroweak
  %baryogenesis,''
  Nucl.\ Phys.\ B {\bf 679}, 246 (2004)
  [arXiv:hep-ph/0309291].
  %%CITATION = HEP-PH 0309291;%%

\bibitem{Moreno:1998bq}
  J.~M.~Moreno, M.~Quiros and M.~Seco,
  %``Bubbles in the supersymmetric standard model,''
  Nucl.\ Phys.\ B {\bf 526} (1998) 489
  [arXiv:hep-ph/9801272].

  \bibitem{singlet}
  See e.g.  M.~Pietroni,
  %``The Electroweak phase transition in a nonminimal supersymmetric model,''
  Nucl.\ Phys.\  B {\bf 402} (1993) 27
  [arXiv:hep-ph/9207227]
  and  S.~Profumo, M.~J.~Ramsey-Musolf and G.~Shaughnessy,
  %``Singlet Higgs Phenomenology and the Electroweak Phase Transition,''
  JHEP {\bf 0708} (2007) 010
  [arXiv:0705.2425 [hep-ph]] and references therein.
  %%CITATION = JHEPA,0708,010;%%
  
 
\bibitem{Fok:2008yg}
  See e.g. R.~Fok and G.~D.~Kribs,
  %``Four Generations, the Electroweak Phase Transition, and Supersymmetry,''
  Phys.\ Rev.\  D {\bf 78}, 075023 (2008)
  [arXiv:0803.4207 [hep-ph]] and references therein.
  
  \bibitem{Shu:2006mm}
  See e.g. J.~Shu, T.~M.~P.~Tait and C.~E.~M.~Wagner,
  %``Baryogenesis from an earlier phase transition,''
  Phys.\ Rev.\  D {\bf 75}, 063510 (2007)
  [arXiv:hep-ph/0610375] and references therein.
  %%CITATION = PHRVA,D75,063510;%%
  
   \bibitem{mjrminprep}
  H.~Patel and M.~J.~Ramsey-Musolf, in preparation.
  
  \bibitem{su2}
   P.~Fileviez Perez, T.~Han, G.~y.~Huang, T.~Li and K.~Wang,
  %``Neutrino Masses and the LHC: Testing Type II Seesaw,''
  Phys.\ Rev.\  D {\bf 78} (2008) 015018
  [arXiv:0805.3536 [hep-ph]];
  %%CITATION = PHRVA,D78,015018;%%
  See also J.~Kehayias and S.~Profumo,
  %``Semi-Analytic Calculation of the Gravitational Wave Signal From the
  %Electroweak Phase Transition for General Quartic Scalar Effective
  %Potentials,''
  arXiv:0911.0687 [hep-ph].
  %%CITATION = ARXIV:0911.0687;%%


 \bibitem{Lyth:1995ka}
  D.~H.~Lyth and E.~D.~Stewart,
  %``Thermal Inflation And The Moduli Problem,''
  Phys.\ Rev.\  D {\bf 53}, 1784 (1996)
  [arXiv:hep-ph/9510204].
  %%CITATION = PHRVA,D53,1784;%%

\bibitem{Knox:1992iy}
  L.~Knox and M.~S.~Turner,
  %``Inflation at the electroweak scale,''
  Phys.\ Rev.\ Lett.\  {\bf 70} (1993) 371
  [arXiv:astro-ph/9209006].
  %%CITATION = PRLTA,70,371;%%

  \bibitem{Nardini:2007me}
  G.~Nardini, M.~Quiros and A.~Wulzer,
  %``A Confining Strong First-Order Electroweak Phase Transition,''
  JHEP {\bf 0709} (2007) 077
  [arXiv:0706.3388 [hep-ph]].
  %%CITATION = JHEPA,0709,077;%%

\bibitem{Pospelov:2006sc}
  M.~Pospelov,
  %``Particle physics catalysis of thermal big bang nucleosynthesis,''
  Phys.\ Rev.\ Lett.\  {\bf 98}, 231301 (2007)
  [arXiv:hep-ph/0605215].
  %%CITATION = PRLTA,98,231301;%%




\bibitem{Cirigliano:2006dg}
  V.~Cirigliano, S.~Profumo and M.~J.~Ramsey-Musolf,
  %``Baryogenesis, electric dipole moments and dark matter in the MSSM,''
  JHEP {\bf 0607} (2006) 002
  [arXiv:hep-ph/0603246].
  %%CITATION = JHEPA,0607,002;%%

\bibitem{maxpaper}
 C.~Wainwright and S.~Profumo,
  %``The impact of a strongly first-order phase transition on the abundance of
  %thermal relics,''
  arXiv:0909.1317 [hep-ph].
  %%CITATION = ARXIV:0909.1317;%%

\bibitem{bdecay1}
 D.~Eriksson, F.~Mahmoudi and O.~Stal,
  %``Charged Higgs bosons in Minimal Supersymmetry: Updated constraints and
  %experimental prospects,''
  JHEP {\bf 0811}, 035 (2008)
  [arXiv:0808.3551 [hep-ph]].
  %%CITATION = JHEPA,0811,035;%%
  
\bibitem{bdecay2}
 M.~Antonelli {\it et al.},
  %``Flavor Physics in the Quark Sector,''
  arXiv:0907.5386 [hep-ph].
  %%CITATION = ARXIV:0907.5386;%%
  
\end{thebibliography}
\end{document}